\documentclass[12pt]{revtex4}
\usepackage{url}

\begin{document}

\title{Notes on Leibniz thought experiment}
\author{Markos Maniatis}
\affiliation{
Departamento de Ciencias B\'{a}sicas,
Universidad del B\'{i}o-B\'{i}o\\
Avda. Andr\'{e}s Bello s/n,
Casilla 447
Chill\'{a}n, 
Chile\\
email: mmaniatis@ubiobio.cl
}


\begin{abstract}
Leibniz thought experiment
of perception, sensing, and thinking is
reconsidered.
We try to understand Leibniz picture
in view of our knowledge of 
basic neuroscience.
In particular we can see how
the emergence of {\em consciousness}
could in principle be understood.
\end{abstract}

\maketitle

Gottfried Wilhelm Leibniz, {\em Monadology} (1714), \S{17}~\cite{Leibniz}:\\
``{\em
Besides, it must be admitted that perception, and anything that depends on it, 
cannot be explained in terms of mechanistic causation -- that is, 
in terms of shapes and motions. 
Let us pretend that there was a machine, which was constructed in 
such a way as to give rise to thinking, sensing, and having perceptions. 
You could imagine it expanded in size (while retaining the same proportions), 
so that you could go inside it, like going into a mill. 
On this assumption, your tour inside it would show you 
the working parts pushing each other, but never anything which 
would explain a perception. So perception is to be sought, 
not in compounds (or machines), but in simple substances. 
Furthermore, there is nothing to be found in simple substances, 
apart from perceptions and their changes. 
Again, all the internal actions of simple 
substances can consist in nothing other than perceptions and their changes.''
}\\

We would like to reconsider Leibniz thought experiment,
presented about three centuries ago.
Today we know, that at a {\em tour inside}
the elementary {\em working parts} are
the neurons, which {\em push each other} by means of
electrical activity.
Following Leibniz closely it is evident that
we can not find at any special location
anything from 
which we could explain perception, sensing or
thinking. This hypothetical special location of perception, sensing
or thinking we would call {\em consciousness}
or {\em I} or {\em self}. However with that notion we would not gain any insight.\\

In contrast, it is immediately clear that any special location of
perception, sensing and thinking leads to contradictions:
suppose, we could detect a special location of perception,
sensing and thinking, then, in a further expansion
in size we could again go inside and would only find
{\em working parts}, {\em pushing each other}.
Indeed, in terms of neurons, 
we know that the neuronal signals
do not converge anywhere. If we consider
for example a visual sense, we know that the signals 
already behind the retina are divided into
different neuronal structures and do not come together anywhere.
Any kind of convergence at some location of
perception would require some new kind of
{\em inner eye}.
We would not be any step forward
-- this 
contradiction is usually denoted as infinite regress (see
for instance the discussion in~\cite{Rosenthal86}).
The crucial point in Leibniz thought experiment
is to recognize that there is no special location of perception,
sensing or thinking. However, in contrast to Leibniz conclusion,
it appears quite natural to explain these
phenomena from the interactions of neurons themselves.
Hence, we should assume that perceptions, sensing, and 
thinking do not arise at a special location,
but are developed in the whole neuronal system.\\

From the picture of working {\em parts pushing each other} 
we draw the conclusion that the processes 
in our brain are deterministic: in any kind of
{\em mill}, consisting of mechanical parts any movement
is caused by the preceding ones. The cascade of
mechanical processes appears to be inescapable.
Indeed {\em free will} has no meaning in
a neurological context since causal processes
contradict anything {\em free}. Also, trying
to employ quantum mechanics to escape from causality~\cite{Bohm,Pribram,QM},
we do not see any way to explain {\em freeness} in
terms of the randomness we encounter in quantum mechanics.\\

Realizing that perception, sensing, and thinking
appear from the cascade of neuronal processes in
an unfree manner we will in the following talk about the
{\em emergence} of these phenomena.
Using the expression {\em emergence} we stress 
that there is nothing like an illusionary ``inner 
location'' where perception, sensing, and thinking
are formed.\\

Let us think this thought further.
Imagine, under anesthetic, in our
brain one neuron after the other would
be replaced by an exact copy.
Since no neuron would be the special location
of perception, sensing or thinking,
we would in no step replace this special location.
After recovering from anesthesia {\em we}
would not recognize any change. The neurons
would interact in the same manner as before
and our perception, sensing, and thinking
would appear in the same way.\\

Of course, it makes no difference whether
we replace the neurons one by one,
or at once.
Likewise in the latter we would develop
perception, sensing, and thinking in
the copy in the same way!
Hence, suppose that we replace under anesthetic
our body by a copy, nothing like {\em I} or
{\em consciousness} or {\em self} would be lost. That is, our perception,
sensing, and thinking is not attached
to certain neurons, but appear from
their activities.

Let us further imagine that we could replace
each neuron in turn by an electronic device, which
replicates exactly the same functionality
as the original neuron.
As before we would not remove in any step
a location of perception, sensing or thinking.
In this way we finally would be replaced by a machine
under anesthetic and this machine would
develop the same perception, sensing and
thinking and {\em we} could not feel any difference!\\

Obviously this picture of the emergence
of perception, sensing, and thinking
is contrary to the accepted opinion.
We are convinced to have some kind
of {\em I} -- a location where perception,
sensing, and thinking is formed.
Why are we subject to this illusion?\\

The crucial point here is to see how the
usage of {\em I} appears in our thoughts:
let us consider an example of a perception,
for instance the smell of an apple.
If we communicate to someone this perception,
we say, for instance: ``I smell the scent
of a fresh apple''. We 
would use grammatical first-person in order
to communicate our own perception,
distinguishing it from a perception
of someone else.
But what happens if we do not communicate
this statement but only realize the smell?
This thinking must be something emergent,
so we can understand it if we suppose that
thinking is nothing but silent
communication.
Hence, we think, in an emergent sense, ``I smell the scent
of a fresh apple''. Therefore we suppose that thoughts are
in this sense always a way of
communication and this implicitly seems 
to originate from a location of perception, likewise
denoted by ``I'' in our example.
In this way the illusion of a
location of perception, sensing,
and thinking is unavoidable --
a machine would develop
the same illusion of a location
of perception, sensing, and thinking
(compare with the ``zombie'' in~\cite{Chalmers97}).\\

The question arises whether we, replaced by a machine, could become immortal?
We guess the answer is yes, supposing we could exactly represent the circuit
of tens of billions of neurons (for an attempt 
see for instance~\cite{bluebrain}).

%
%
%

\end{document}